\documentclass[fleqn,11pt]{article}
\usepackage{graphicx}
\usepackage{amsmath}
\usepackage{enumerate}

\raggedright

\begin{document}

\noindent{\LARGE \textbf{Turing pattern with proportion preservation}}
\vspace{5mm}

\noindent Shuji Ishihara$^{1}$ and Kunihiko Kaneko$^{1,2}$
\vspace{5mm}

\noindent 
$^1$Department of Pure and Applied Sciences, University of Tokyo,
Komaba, Meguro-ku, Tokyo 153-8902, Japan \\ $^2$ ERATO Complex Systems
Biology Project, JST\\ 
\vspace{5mm}

\noindent Correspondence should be addressed to S. Ishihara\\[-2mm]
\noindent E-mail : shuji@complex.c.u-tokyo.ac.jp \\[-2mm]
\noindent Department of Pure and Applied Sciences, College of Arts and Sciences,\\[-2mm]
\noindent The University of Tokyo, 3-8-1 Komaba, Meguro-ku, Tokyo 153-8902, Japan.\\[-2mm]
\noindent Tel/Fax : +81-3-5454-6731\\
\vspace{5mm}

Although Turing pattern is one of the most universal mechanisms for
pattern formation, in its standard model the number of stripes changes
with the system size, since the wavelength of the pattern is
invariant: It fails to preserve the proportionality of the pattern,
i.e., the ratio of the wavelength to the size, that is often required
in biological morphogeneis.  To get over this problem, we show that
the Turing pattern can preserve proportionality by introducing a
catalytic chemical whose concentration depends on the system size.
Several plausible mechanisms for such size dependence of the
concentration are discussed.  Following this general discussion, two
models are studied in which arising Turing patterns indeed preserve
the proportionality.  Relevance of the present mechanism to biological
morphogenesis is discussed from the viewpoint of its generality,
robustness, and evolutionary accessibility.

\vspace{2mm}


\newpage
\parskip=3.5mm
\parindent=5mm
\section{Introduction \label{sec:intro}}

~~~~~Since the seminal paper by Turing (1952), a large variety of
pattern formation phenomena in nature has been explained by his
theory. The original motivation of Turing himself lied in the
explanation of biological morphogenesis, as was succeeded to Gierer
and Meinhardt (Gierer and Meinhardt, 1972), and others over a half
century (Gray and Scott, 1984; Murray 1993; Pearson, 1993; Kondo and
Asai, 1995; Meinhardt and Gierer, 2000). Although the Turing pattern
is one of the most beautiful and ubiquitous mechanisms for
morphogenesis, frequent criticism raised to it is non-adjustability of
the characteristic wave length of the pattern against the system size.
Because the generation of the Turing pattern comes from instability of
an uniform state over a certain range of wavelengths, the possible
range of the wavelengths is pre-fixed, and is invariant against the
change of the system size. Hence the number of segments or stripes is
proportional to the system size as shown in Fig.~\ref{fig:Prop}~(a).
In contrast, the number of segments or stripes, rather than the
wavelength, is often invariant against the change of the size in many
biological systems as shown in Fig.~\ref{fig:Prop}~(b). In a
biological system, the scale of pattern is often proportional to the
system size, and also it is desirable to have such proportionality in
many situations. However, neither the Turing pattern nor the
positional information theory by Wolpert (1969) which assumes the
morphogen gradient to convey positional information satisfies the
scale-invariance.

For example, it is observed that the patterns in \textit{Hydra} and
\textit{Dictyostelium discoideum} slugs have proportionality with the 
size. In \textit{Dictyostelium discoideum} the ratio of two different
cell types is almost fixed, independent of the size. Transgenic mice
indeed preserve the body proportion despite their larger size in
phenotype (Palmiter et al., 1982). In the development of
\textit{Drosophila Melanogaster}, it was reported that the expression
of gap gene \textit{hb} is robust and preserves the proportion over
different sizes of individual eggs (Houchmandzadeh et al., 2002). The
proportion preservation is important for robust morphogenesis in
general.

In the present paper we discuss a general mechanism which enables the
proportionality of wavelength with the size in Turing patterns as in
Fig.~\ref{fig:Prop}~(b).

To explain the proportion regulation of different cell types for
\textit{Dictyostelium discoideum} slug,  Meinhardt (1982) proposed an 
activator-inhibitor model in which the ratio of two cell types is
preserved.  The mechanism works well for a pattern with only a single
boundary between two types, but is not valid for multiple stripes
pattern.  Theory based on globally coupled dynamical systems also
provides a regulation mechanism for the ratio of different cell types
(Kaneko and Yomo, 1994, 1999; Mizuguchi and Sano, 1995; Furusawa and
Kaneko, 2001), but a proportion preservation of a pattern with
multiple stripes is not discussed yet.

As an extension of Turing pattern, Othmer and Pate (1980) showed that
if diffusion constants depend on the concentration of auxiliary
chemical factor which has size-dependence, then size-invariant pattern
formation is possible, where the size-dependence of the auxiliary
chemical concentration is provided by choosing a proper boundary
condition (see also Pate and Othmer, 1984; Dillon et al.,
1994). Hunding and S\o rensen (1988) also discussed a simple mechanism
to explain such concentration-dependent diffusion by an auxiliary
chemical factor. In these models, it is necessary that all the
diffusion coefficients are regulated in the same manner. Apart from
Turing pattern, a model for the proportion regulation in
\textit{Drosophila Melanogaster} was proposed by Aegerter-Wilmsen et al.
(2005), also by assuming effectively concentration-dependent
diffusion.  It is not yet sure if such regulation of diffusion is
really adopted to control the proportionality.

In this paper we discuss mechanisms of scale-invariant Turing pattern
without considering any variations of diffusion constants. Instead we
seek for a possibility that concentration of some chemical changes
with some power of the system size, which influences the rate of
reaction for the Turing pattern so that the size-invariance is
generated.  We introduce a chemical component whose concentration
depends on the size of the system. In Section \ref{sec:sdc} we discuss
several possibilities for such size-dependent concentration of a
chemical.  Following this general discussion, we give two specific
examples leading to the Turing pattern whose wavelength is
proportional to the system size. The first example introduced in
Section \ref{sec:Model1} adopts a size-dependent auxiliary chemical
component, in the same way as Othmer and Pate (1980) or Hunding and
S\o rensen (1988), while we believe it is simpler than the earlier
models, and is also plausible biologically because change between
active and inactive forms adopted therein is ubiquitous in a
biochemical process. The second example introduced in Section
\ref{sec:Model2} contains only two chemical components, which is the
minimum number for the Turing instability. The model provides a novel
mechanism for the proportionality preservation based on the
conservation of some quantity.  Due to the size dependence of the
conserved quantity, the scale-invariance in Turing pattern is
resulted.  In Section \ref{sec:discussion} we summarize the mechanisms
for the proportion preservation and discuss possible relevance of them
to biological morphogenesis.

Before discussing the mechanisms, we make one remark what we have in
mind with the term ``system'' in the present paper.  In a class of
examples, the system refers to a cell, in which case the boundary of
the system is a membrane, while in some other cases, the system refers
to cell aggregates (or tissue).  As a mathematical expression of
reaction-diffusion equation the two cases are treated in the same way,
and thus we discuss the two cases together by adopting the term
``system''.

\section{Size dependent concentration \label{sec:sdc}}

~~~~~Following Section \ref{sec:intro}, we seek for a mechanism in
which concentration of some chemical components changes with the
system size.  Let us consider the case in which a single component W
satisfies the size-dependence in concentration. Here we will discuss
several possible mechanisms in which the concentration of W indeed
changes with some power of the system size.  We take a
three-dimensional system with a size scale $\sim L$, and assume that
the size of the system (e.g., a cell) varies keeping conformity in
shape, so that the volume of the system is proportional to $L^3$,
while the area of the boundary increases with $L^2$.  We further
assume that the diffusion of W is so rapid as $\sqrt{D_w/\gamma_w} >
L$ where $D_w$ and $\gamma_w$ are the diffusion coefficient and the
degradation rate of W respectively, so that W is distributed almost
homogeneously. This condition, however, is not so restrictive to
realize the size-dependence, and the argument below can be generalized
even by relaxing the condition.

\begin{enumerate}[\bfseries (A)]  

\item  
Consider the case in which the total quantity of W is conserved
against the change of the system size (or through the growth), as
shown in Fig. \ref{fig:turingcase} (A). In this case the concentration
of W is proportional to $L^{-3}$ due to the conservation of the total
amount and the dilution by the increase of the size.  Here the case
with only a single chemical component (W) is discussed, but $W$ can be
a sum of multiple components if the total sum of them is conserved.
In Section \ref{sec:Model2}, we discuss an example of such case, which
results in the size-invariant Turing pattern.
 
\item 
Consider the case in which W is generated whole through the system
at a rate $g$, while W escapes out of the system only through the
boundary, as in Fig. \ref{fig:turingcase} (B-(i)).  As another
possibility, consider the case in which W is decomposed by enzymes bounded on the
membrane (if a system is a cell) or by specific cells that are located
at the boundary of a tissue as in Fig. \ref{fig:turingcase} (B-(ii)).

The case (i) is the same as that discussed by Othmer and Pate
(1980). In this case, the boundary condition is represented by the
following equation
\begin{eqnarray}
  -\vec{n}\cdot D_w \nabla w = b w \label{eq:bound}
\end{eqnarray}
where $\vec{n}$ is an unit vector perpendicular to the boundary, and
$b$ is the mass transfer coefficient of W at the surface of the
system.  In the cases (ii), degradation of W is catalyzed only at the
boundary.  Then $w(\vec{r},t)$ follows the equation
\begin{eqnarray}
  \frac{\partial w}{\partial t} = D_w \triangle w + g - \gamma w
  \delta(\vec{r}-\vec{r}_s), \label{eq:wB}
\end{eqnarray}
where $\vec{r}_s$ denotes the coordinate of the boundary. 

In both cases, W is distributed almost homogeneously in the system if
the diffusion coefficient is sufficiently large. In the steady state,
concentration $w$ in the system is evaluated by the integration, where
W is synthesized with the rate proportional to $L^3$ and is
decomposed in proportion to $L^2$. Thus the abundances of W are
proportional to $L$.

It is often the case that the generation of some chemical factors are
limited at a localized region in a system. Bcd-protein in the embryo of
\textit{Drosophila} is an example, where Bcd-mRNA is fixed in 
the anterior of the cell. In such case, the rate of synthesis of W is
independent of the system size ($L^0$), and thus the concentration of W is
proportional to $L^{-2}$.

\item
As is shown in Fig. \ref{fig:turingcase} (C), W flows into a system
from (or is synthesized by a chemical factor from) the environment of
the system, and is decomposed within the system. Then the former rate
is proportional to $L^2$, and the latter to $L^3$, so that the
concentration of W is proportional to $L^{-1}$. This situation is
typical for morphogenesis where each part of the embryo transmits and
receives signals with each other. As another example, cAMP in a cell
is synthesized by the membrane-bound enzyme Adenylcyclase, and thus
the concentration of cAMP follows the above scaling relation.

\item
Consider a chemical factor Z that is synthesized whole through the
system while it is degraded on the boundary. Then the concentration of Z,
$z$, is proportional to the system size ($z \propto L$). Also consider
a chemical W that is synthesized whole through the system while it is
degraded, catalyzed by two molecules of Z, such as $2Z + W \to 2Z +
G$, as shown in Fig.~\ref{fig:turingcase} (D). Then the concentration
of W is proportional to $ L^{-2}$.  In general, cooperative reactions
as in this example can induce various $L$ dependence.
\end{enumerate}

Of course, some other situations are possible in which $w$ depends on
the size of a system.  Next we give specific examples of
reaction-diffusion equations that leads to the scale-invariant Turing
pattern, based on this scaling behavior of a chemical W.  In these
models, chemical factors U and V regulate each other, which, we
assume, are impenetrable through the boundary (membrane).

\section{Model~I : Turing model with a size regulator\label{sec:Model1}}
\subsection{Controlling proportionality}
~~~~~Here we give an example of size-invariant Turing pattern, based
on the chemical $W$ with the size-dependent concentration in Section
\ref{sec:sdc}.  Consider a reaction-diffusion system composed of three
chemical components U, V, W. The concentrations of U and V at time $t$
and at position $\vec{r}$, $u(\vec{r},t)$ and $v(\vec{r},t)$, obey the
following equations
\begin{subequations}
\begin{eqnarray}
  \frac{\partial u}{\partial t} & = & D_u \triangle u +f(u,v;w) \label{eq:Turinga}\\
  \frac{\partial v}{\partial t} & = & D_v \triangle v +g(u,v;w). \label{eq:Turingb}
\end{eqnarray}
\end{subequations}
W is a factor controlling the reactions, which is a size-dependent
component at the same time. The reaction terms are represented by
$f(u,v;w)$ and $g(u,v;w)$, and the wavelength of U-V pattern is
controlled by the concentration of W ($w$). In general, the change of
$w$ is accompanied with the change of the homogeneous steady state
itself, and as a result the characteristic wave length at the unstable
uniform state may change in a complicated manner.  Here, we just give
two simple classes of reaction equations that satisfy the
scale-invariant pattern formation.

In the first case, all the reactions for U and V are homogeneously
regulated by W, i.e. $f(u,v) \propto w^{\mu}$ and $g(u,v) \propto
w^{\mu}$. Here, by the spatial scale transformation $x \to
x/w^{\mu/2}$, the $w$-independent differential equations are obtained
for a steady-state pattern.  Such regulation was also assumed in
earlier study by Saunders and Ho (1995), but it might not be so
natural, as W regulates all the reaction process in the same manner.

As far as we know, the second case has been slipped over, in
which the reactions for U and V have the functional forms
\begin{eqnarray}
f(u,v;w) = F(w^{\nu}u,w^{\nu}v), ~~~~g(u,v;w) =G(w^{\nu}u,w^{\nu}v). \label{eq:exp}
\end{eqnarray}
In this case a homogeneous fixed point given by the conditions
$f(u,v;w)=g(u,v;w)=0$ is realized at $\hat{u} \equiv
w^{\nu}u=\hat{u}^0$ and $\hat{v} \equiv w^{\nu}v=\hat{v}^0$, where
$(\hat{u}^0,\hat{v}^0)$ is the solution of $F(u,v)=G(u,v)=0$, so that
the linearized partial differentiations around the fixed point are
given by $f_u=w^{\nu} F_{\hat{u}}(\hat{u}^0,\hat{v}^0)$ and so forth.
By the transformation of $(u,v)$ to $(\hat{u},\hat{v})$, the equations
are invariant under the spatial scale transformation $x \to
x/w^{\nu/2}$,

In the above two cases, the characteristic wavelength for the unstable uniform
steady state of U, V is controlled by the concentration of W. In the
next subsection, by taking a simple specific reaction-diffusion model
we show that this latter case arises rather naturally.

\subsection{A reaction-diffusion model}
~~~~~Here we study the following reaction-diffusion model based on
Brusselator (Prigogine and Lefever, 1968; Nicolis and Prigogine,
1977), in addition to the size-regulator W;

\begin{tabular}{clcccc}
\textbf{(\,I\,)}  & U is generated by A at a constant rate &:& A $\stackrel{k_A}{\to}$ U  \\
\textbf{(II)}  & U is activated into U$^*$ by W with a reversible reaction  &:& U $+$ W $\mathop{\rightleftharpoons}_{k^{-1}_U}^{k_U}$ U$^*$ \\
\textbf{(III)} & V is activated into V$^*$ by W with a reversible reaction &:& V $+$ W $\mathop{\rightleftharpoons}_{k^{-1}_V}^{k_V}$ V$^*$ \\
\textbf{(IV)} & U$^*$ changes to V$^*$ at a constant rate &:&  U$^*$ $\stackrel{k_b}{\to}$ V$^*$ \\
\textbf{(V)}  & Dimer of U$^*$ catalyzes V$^*$ into U$^*$ &:& 2U$^*$ + V$^*$ $\stackrel{k_a}{\to}$ 3U$^*$\\
\textbf{(VI)} & U$^*$ is degraded at a constant rate &:&  U$^*$ $\stackrel{k_G}{\to}$ G  
\end{tabular}

\noindent
The model is illustrated in Fig.~\ref{fig:Brusselator}.  In the model,
U and V have active and inactive states, and can react only in its active
state ``*''.  At the same time, U and V are activated by W.

Under a proper rescaling and redefinition of the parameters, the
rate equations for the system are given by
\begin{subequations}
\begin{eqnarray}
  \dot{u}_i &=& A-k_U\,w u_i + k_U^{-1} u_a \\
  \dot{u}_a &=& k_U \,w u_i -k_U^{-1} u_a - u_a - B u_a + u_a^2 v_a\\
  \dot{v}_i &=& - k_V w v_i + k_V^{-1} v_a\\
  \dot{v}_a &=& k_V w v_i - k_V^{-1} v_a + B u_a - u_a^2 v_a
\end{eqnarray}
\end{subequations}
where $u_i,~u_a,~v_i,~v_a$, and $w$ are concentrations of U, U$^*$, V,
V$^*$, and W respectively. Let us assume that the reversible reactions
between active and inactive states (II,~III) are sufficiently rapid
and in equilibrium. Then, the ratio of U to U$^*$ (V to V$^*$) is
given by a constant $u_a = \left(k_U\,w/ k_U^{-1} \right) u_i$ ~($v_a =
\left(k_Vb/ k_V^{-1} \right) v_i $), by which their terms
with $u_a$ and $v_a$ are replaced. As a result, we obtain the
equations of $u \equiv u_a+u_i$ and $v \equiv v_a+v_i$ as:
\begin{subequations}
  \begin{eqnarray}
	\dot{u} & = & A -m(w)\, u + m(w)^2 n(w) \, u^2 v- B m(w)\, u \\
	\dot{v} & =  & -m(w)^2 n(w) \,u^2 v + B m(w)\, u 
  \end{eqnarray}
\end{subequations}
where $m(w) = k_U\,w/(k_U\,w + k_U^{-1}) $ and $n(w) = k_V\,w/( k_V\,w
+ k_V^{-1})$.  Note that $u_a = m(w) u $ and $v_a = n(w) v$.  In the
case $k_U^{-1} \gg k_U $ and $k_V^{-1} \gg k_V $, where inactive
states are dominant, $m(w) \sim (k_U / k_U^{-1}) w$ and $n(w) \sim
(k_V / k_V^{-1}) w$ approximately.  Accordingly, the conditions of
Eq.~(\ref{eq:exp}) are satisfied with $\nu=1$. Notice that this
satisfaction of Eq.~(\ref{eq:exp}) is not specific to this model, but
is general when the chemicals have active and inactive states and only
the former participates in the reaction.

Now we consider the situation given by \textbf{(B)} in
Section \ref{sec:sdc}, where W is synthesized at a limited domain in the
system. We just consider an one-dimensional pattern, and study a model
represented by the following partial differential equations;
\begin{subequations}
  \begin{eqnarray}
	\frac{\partial u}{\partial t } &= & D_U \frac{\partial^2 u}{\partial x^2}  + A -m(w)\, u + m(w)^2 n(w) \, u^2 v- B m(w)\, u \\
	\frac{\partial v}{\partial t } &= & D_V \frac{\partial^2 v}{\partial x^2}  - m(w)^2 n(w) \,u^2 v + B m(w)\, u  \\
	\frac{\partial w}{\partial t } &= & D_W \frac{\partial^2 w}{\partial x^2}  + H(x) - \gamma w \label{eq:PB3}
  \end{eqnarray}
\label{eq:simueq}
\end{subequations}
\hspace{-1.5mm}where $H(x)=1$ for $0<x<x_0$ and 0 otherwise. To represent the
synthesis of W in some definite area of the system irrespective of the
system size, the constant $x_0$ is independent of $L$ and is set at
$1.0$ in the simulation.  The last term $-\gamma w$ represents the
escape (or decomposition) of W through the surface.  We have carried
out simulations on the temporal evolution of $u_n, v_n ~ (n=0 \sim
N-1)$ under Dirichlet boundary condition $U(0)=U(L)=V(0)=V(L)=0.0$.
Since we consider a one-dimensional direction of a three dimensional
system with the size $L$, the flow-out of the escaping chemicals at
the boundary should be proportional to $L^2$, so that the $\gamma$
term in the above model equation should be scaled by $L^{2}$.  Under
these assumptions on $H(x)$ and $\gamma$, the discussion in
\textbf{(B)} of Section \ref{sec:sdc} is valid.  Indeed, we have confirmed
that the concentration $w$ is proportional to $L^{-2}$, in the
simulation for large $D_W$. Thus we take $w=W_0 L^{-2}$ for the
simulations below.

In Fig.~\ref{fig:model1x-L}, we plot the wavelength $\xi$
corresponding to the wavenumber that leads to the largest eigenvalue
in the linear stability analysis, for a given system size $L$. $\xi$
increases in proportion to $L$, and thus the generated pattern from
this instability is expected to preserve the proportion.

The results of the simulation are shown in
Fig.~\ref{fig:SimulationTuring}, which clearly show that the number of
stripes does not change against the change of system size $L$ as long
as it is sufficiently large.  For small size, because $w$ is large,
the saturation in the terms in $m(w)$ (or $n(w)$) is not negligible,
so that the number of stripes is decreased.

\subsection{Simplicity of the mechanism \label{sec:CondTuring}}
~~~~~The above example gives us a simple but plausible model for the
regulation of the wave length, which leads to the 
proportion preservation for a pattern generated by Turing instability.  As shown in
Fig.\ref{fig:TuringMech}, conditions for this proportion preservation
are summarized as follows:
\begin{enumerate}[\sffamily\bfseries (i)]  
\item Each chemical component has active and inactive states.  
The activation is reversible and is catalyzed by a chemical factor W
(whose concentration changes with the size).  Only chemicals in the
active state can participate in reactions.
\item The reaction-diffusion system shows Turing instability.
\item W is generated at a localized region in the system (cell), diffuses rapidly, and
goes out of, or is degraded on, the surface (membrane).
\end{enumerate}
These are the only conditions for the proportion preservation of a
Turing pattern, which works regardless of the specific choice of a
model.  The condition \textbf{\textsf{(iii)}} can be replaced by some
other conditions in which the concentration of the factor is scaled as
$w \propto L^{-2}$. Notice that the above conditions are independent
of each other so that they would be easily satisfied by combining each
process that satisfies each condition. Thus, a size-invariant Turing
pattern by the above conditions may be achieved easily through the
evolution.

\section{Model~II : Model with a conserved quantity\label{sec:Model2}}
\subsection{Size invariant Turing instability by the conserved quantity}
~~~~~Here we give another model with two chemical components, which are
regarded as two states of a single chemical species. At the same time
both U and V are not synthesized or decomposed, so that the total
quantity of the chemical components is conserved. According to the
mechanism \textbf{(A)} discussed in Section \ref{sec:sdc}, we seek for the
possibility of the proportion preservation in  this model. For simplicity,
we assume one-dimensional system in this section.

Consider two components U and V that regulate the concentration of each other
through a reaction, as shown in Fig.~\ref{Fig:Model2}. Then the reaction-diffusion 
equations are represented by
\begin{subequations}
  \begin{eqnarray}
	\frac{\partial u}{\partial t} &=& D_u \frac{\partial^2 u}{\partial x^2} + F(u,v) \\
	\frac{\partial v}{\partial t} &=& D_v \frac{\partial^2 v}{\partial x^2} - F(u,v) 
  \end{eqnarray}
\label{eqn:rd}
\end{subequations}
Hence the total quantity of U and V is conserved;
\begin{eqnarray}
  S \equiv \int dx \left( u+v \right) = \mbox{constant}.
\end{eqnarray}
As discussed in the case \textbf{(A)} in Section \ref{sec:sdc}, the increase in
the system size leads to the dilution of the concentration of $S$.

In the steady homogeneous state, the Jacobian matrix for the reaction
terms is given by
\begin{eqnarray}
  J=\left(
  \begin{array}{cc}
	 F_u  &   F_v \\
	-F_u  &  -F_v  
  \end{array}
\right)
\end{eqnarray}
where $F_u$ denotes the partial derivative of $F$ by $u$ at a
homogeneous steady state of Eq.~(\ref{eqn:rd}), and so forth. Through
the stability analysis of the Fourier transform of the linearized
equation around the homogeneous state by using $J$, the wave number
that has the largest eigenvalue is obtained, which gives the wavenumber
that grows most rapidly from this unstable homogeneous state.  This
wavenumber is given by
\begin{eqnarray}
  k_m^2 = \frac{-\hat{D}(F_u+F_v)+ (1+\hat{D}) \sqrt{F_u F_v \hat{D}}}{\hat{D} (\hat{D}-1)}
\label{eq:Awacond}
\end{eqnarray}
with $\hat{D}\equiv D_v/D_u$, where $\hat{D}$ is larger than $1$ for
Turing instability to occur (Turing, 1952). To preserve the
proportional pattern by increasing the length of the system $L$, it is
necessary that both $F_u$ and $F_v$ behave so as to $k_m$ scales as
$L^{-1}$ in Eq.~(\ref{eq:Awacond}), at least approximately.  Below, we
give an explicit example corresponding to this case.

\subsection{An explicit reaction-diffusion model with a conserved quantity}
~~~~~Consider the following  reaction diffusion system corresponding to the
reactions shown in Fig.~\ref{Fig:Model2}.
\begin{subequations}
  \begin{eqnarray} 
	\frac{\partial u}{\partial t} &=& D_u  \frac{\partial^2 u}{\partial x^2} + u^3v - Bu^2 \\ 
	\frac{\partial  v}{\partial t} &=& D_v \frac{\partial^2 v}{\partial x^2} - u^3v +  Bu^2 \end{eqnarray}
\label{eq:simueq2}
\end{subequations} 
This system is a modified version of the Brusselator, so that supplies
and degradations of substances are excluded (Awazu and Kaneko,
2004). Although the reaction term $u^3v$ is higher than the original
Brusselator, indeed, the reaction with a lower order cannot satisfy
the requirement of the last subsection for the proportion
preservation. As far as we have examined, this choice is one of the
simplest to satisfy the requirement (See Appendix A).  Also, in a
biological system, such high order catalysis is not so uncommon. Hence
we adopt this reaction model.

In this system, the corresponding homogeneous fixed point $(u_0,v_0)$
is given by $u_0v_0=B$ and $u_0+v_0=S/L$.  Note that $u_0$ is almost
proportional to $S$ as long as $S$ is sufficiently large.  Jacobian
around the uniform steady state is given by
\begin{eqnarray}
J=\left( \begin{array}{cc} Bu_0& {u_0}^3 \\ -Bu_0& -{u_0}^3
\end{array}
\right)
\end{eqnarray}
If $\hat{D}$ is large enough, the dominant term in
Eq.~(\ref{eq:Awacond}) is the one containing $\sqrt{F_u F_v}$, because
other terms are lower order with regards to $\hat{D}$.  Thus, $k_m^2
\sim \sqrt{{u_0}^4} \sim S^2$ holds in the model, which results 
in the proportion preservation.  

We plot the characteristic wavelength $\xi=2\pi/k_m$, corresponding to
the most unstable mode given by the linear analysis of $J$, for the
system size $L$ in Fig.~\ref{fig:model2anal}. $\xi$ increases in
proportion to $L$ over a wide range of $L$ for sufficiently large $S$,
thus leading to a size-invariant pattern.  We have carried out
numerical simulation of Eq.~(\ref{eq:simueq2}).  The results are shown
in Fig.~\ref{fig:SimulationTuring2}, which support the above
estimation to realize the size-invariant Turing pattern formation.

The model we give here is one of the simplest, in the sense that it
contains the lowest order polynomial reaction term among such
equations, as is also explained in the Appendix A, where more detailed
estimations as well as some other equations leading to the
size-invariant pattern formation.

\section{Summary and Discussion \label{sec:discussion}}

~~~~~In this paper, we have discussed a possible mechanism of
proportion regulation based on the control of the reaction rate in
reaction-diffusion systems.  We have introduced a morphogen W which
itself does not convey positional information (Wolpert, 1969), but
works as a carrier of information on the size of the system.  It is
important to recognize that the proportion preservation is possible by
such simple mechanism.  We have discussed several possible schemes
that can naturally realize the proposed mechanism in a biochemical
system.

In some earlier studies and in our model, it is assumed that there are
chemical factors whose concentration depends on the system size. As
discussed by Hunding and S\o rensen (1988), a candidate of such
chemical component is cAMP, which is synthesized by the membrane-bound
enzyme Adenylcyclase. Indeed, cAMP is involved in a number of
important biological processes.

Some proteins can also fit as the size regulator W. A candidate is the
product of the gene \textit{staufen} (\textit{stau}) working in the
early development of \textit{Drosophila}.  Houchmandzadeh et
al. (2002) reported that in the \textit{Drosophila} embryo, the domain
boundary of zygotic gene \textit{hunchback} (\textit{hb}) expression
is tuned precisely at a half of the embryo, despite individually
fluctuating embryo size and expression of its direct regulator
Bicoid. It is discussed that maternal gene
\textit{stau} may play a major role to control such positioning of
\textit{hb} expression, as mutants lacking \textit{stau} lose precise
expression boundary of
\textit{hb} at the half of the embryo. Although a model based on the
effective change of diffusion constant was proposed by
Aegerter-Wilmsen et al. (2005) recently, it may be interesting to seek
for the possibility that \textit{stau} may work as a size-regulator W
with the mechanism (A) or (B) in Section \ref{sec:sdc}.  In general,
it will be interesting to search for some molecules that work as size
regulators, or carry out a knock-out experiment on a candidate
molecule of such regulator.

In Section \ref{sec:Model1}, we have introduced a simple model in
which the Turing pattern is size-invariant. The proposed scheme for
the proportion preservation there is rather general and robust, and at
the same time is naturally realized in a biological system. We give
three conditions for the scheme, which are rather simple and plausible
to be achieved in biological morphogenesis. Additionally these
conditions are independent of each other, which is a good feature from
an evolutionary viewpoint, because they can be established one by one
through evolution, without any influence with each other.  Consider
two neighbor species with similar proportional organization, but with
different sizes. Most of the genes are common between the two, and
they may share the same diffusion coefficients for most of their
products. Under these conditions, ordinary Turing pattern or other
mechanisms cannot explain conformity in their morphology for two
species with different sizes.  On the other hand, in the mechanism we
proposed in Sec. \ref{sec:Model1}, control of solely a single chemical
W can lead to proper adaptive patterning. This is one of evolutionary
advantages of the present mechanism.

Independence of each condition is also a good feature for an
experimetal reailzation of the present mechansim.  One can realize the
present proportion-prserving Turing pattern based on the established
experimental examples (Castets et al., 1990; Ouyang and Swinney
1990). Because ordinary Turing pattern has just one definite
wavenumber, it is interesting experimetally to construct a pattern
whose intrinsic scale is changed flexibly according to environmental
conditions, using the mechanism discussed in this paper.

In Section \ref{sec:Model2}, we have given an example for the
size-invariant Turing pattern realized by chemical reactions with a
conserved quantity as proposed in (\textbf{(A)} in Section
\ref{sec:sdc}). Existence of such conserved quantity in morphogenesis
may be a rather natural assumption.  However, in contrast to the
models (conditions) given in Section \ref{sec:Model1}, the model here
may lack generality, since the condition for the scale-invariance,
i.e., the combination of exponents, may be rather specific. Also,
accurate control for the initial value of the conserved quantity $S$
(the total quantity of U and V) may be required.  In this sense,
search for the mechanism in Section \ref{sec:Model1} may be more
important in a biological context.

As another possible explanation for the size-invarint pattern, one
could assume that Turing mechanism works only in certain period of
early development, leading to a pattern with differentiated cell
types, and then the cells grow at the same rate, keeping the
proportionality.  In general, this mechanism has a low tolerance for
the individual fluctuation of body size at the earlier stage of
development (Houchmandzadeh et al., 2002). Also the growth process
keeping the proportionality is required, and the mechanism is
vulnerable by disturbance through the development.  On the other hand,
in our mechanism the proportionality is tolerant against size
fluctuations, and is recoverable again such disturbances or external
manipulation.

At last we give a speculation on a relationship between the size
regulation and pattern formation. In general the size of an organ is
much flexibly regulated by the mutual compensation between cell size
and cell number (Frabkenhauser, 1945; Potter and Xu, 2001).  However,
the mechanism of size control in development is not so clear
yet. Potter and Xu (2001) discuss the relationship between size
regulation and pattern formation, in which mutations in genes
regulating pattern result in the changes in total tissue mass. If the
size regulator W discussed in this paper also concerns with size
control through its concentration, the pattern formation process is
tightly coupled with the organ size. It may be interesting to seek for
this possibility, since the present mechanism then allows for adaptive
control of the pattern scale as well as the organ size.

In conclusion, we have shown that morphogenesis with proportion
preservation is possible under Turing instability, by simply utilizing
a catalytic molecule whose concentration is properly scaled with the
system size.

\section*{Appendix A : On the Equations (\ref{eq:simueq2})}
~~~~~Here we explain why we choose Eq.~(\ref{eq:simueq2}) as a model to
satisfy $k_m \sim L^{-1}$.  First consider the reaction-diffusion
equation in a polynomial form
\begin{subequations}
  \begin{eqnarray}
	\frac{\partial u}{\partial t} &=& D_u \frac{\partial^2 u}{\partial x^2} + u^mv^n - Bu^l \\
	\frac{\partial v}{\partial t} &=& D_v \frac{\partial^2 v}{\partial x^2} - u^mv^n + Bu^l
  \end{eqnarray}
\end{subequations}
Then the steady uniform solution $(u_0,v_0)$ is given by
\begin{subequations}  
\begin{eqnarray}
  u_0^{m-l}v_0^n&=&B \label{eq:uvst1}\\
  u_0+v_0&=&S/L \label{eq:uvst2}
\end{eqnarray}
\label{eq:uvsteady}
\end{subequations}
and the Jacobian at this solution is given by
\begin{eqnarray}
J=\left(
\begin{array}{cc}
  (m-l) B u_0^{l-1} & n Bu_0^l v_0^{-1}\\
  -(m-l) B u_0^{l-1} &  -n Bu_0^l v_0^{-1}
\end{array}
\right)  
\end{eqnarray}
The steady state solution $(u_0,v_0)$ is represented as crossing
points of Eq.~(\ref{eq:uvst1}) and (\ref{eq:uvst2}).  When $m>l$, the
relationship is represented as in Fig. \ref{fig:uvApend1} on the
$u$-$v$ plane. Then the crossing point satisfies $u_0 \propto S/L$
approximately for large $S$. On the other hand, if $m<l$, $v_0$ is an
increasing function of $u_0$ in the relationship
Eq.~(\ref{eq:uvst1}). Then the crossing point does not satisfy $u_0
\propto S/L$. Hence we assume $m>l$ here.
Recall that if $D_v/D_u$ is sufficiently large, the leading term
determining the characteristic scale length $\xi$ is given by
$\sqrt{-F_vG_u}$ ~(Eq.~(\ref{eq:Awacond})),
\begin{eqnarray}
 \xi^{-2} = k_m^{2} \sim \left( u_0^{2l-1} v_0^{-1} \right)^{1/2} \sim
 u_0^{\frac{2l-1}{2}+\frac{m-l}{2n}} \sim
 L^{-\frac{2l-1}{2}-\frac{m-l}{2n}}
\end{eqnarray}
(note $v_0 \sim u_0^{(l-m)/n}$ from Eq.~(\ref{eq:uvst1})). Since the
exponent has to be -2 to sustain the proportionality, we get the condition
\begin{eqnarray}
  n=\frac{m-l}{5-2l}
\end{eqnarray}
Suppose that $n,~m,~l$ are positive integers.  Then because $m-l>0$,
$l$ can be only $1$ or $2$.  Now by choosing $m=3, n=1, l=2$
Eq.~(\ref{eq:simueq2}) (Model II) is derived, which is the system with
the lowest degree of exponent.  Another choice will be $m=4, n=1, l=1$
which leads to the following equation:
\begin{subequations}
  \begin{eqnarray}
	\frac{\partial u}{\partial t} &=& D_u \frac{\partial^2 u}{\partial x^2} + u^4v - Bu \\
	\frac{\partial v}{\partial t} &=& D_v \frac{\partial^2 v}{\partial x^2} - u^4v + Bu
  \end{eqnarray}
\label{eq:ModelIIb}
\end{subequations}

\section*{Acknowledgment}
The authors are grateful to A. Awazu, K. Fujimoto, T. Shibata, and
H. Takagi for discussions.


\section*{References}

\noindent Aegerter-Wilmsen, T., Aegerter, C. M., Bisseling, T., 2005.  
Model for the robust establishment of precise proportions in the early
Drosophila embryo. J. Theor. Biol. 234, 13-19

\noindent Awazu, A., Kaneko, K., 2004.  Is relaxation to equilibrium 
hindered by transient dissipative structures in closed systems?
Phys. Rev. Lett. 92, 258302

\noindent Castets, V., Dulos, E., Boissonade, J., and De Kepper, P., 1990. 
Experimental evidence of a sustained standing Turing-type
nonequilibrium chemical pattern.  Phys. Rev. Lett. 64, 2953-2965

\noindent Dillon, R., Maini, P. K., Othmer, H. G., 1994.  Pattern formation 
in generalized Turing systems I. Steady-state patterns in systems with
boundary conditions.  J. Math. Biol. 32, 345-393.

\noindent Frankenhauser, G., 1945. The effects of changes in chromosome 
number on amphibian development. Q. Rev. Biol. 20, 20-78

\noindent Furusawa, C., Kaneko, K., 2001. Theory of Robustness of Irreversible
Differentiation in a Stem Cell System: Chaos Hypothesis.
J. Theor. Biol. 209, 395-416

\noindent Gierer, A., Meinhert, H., 1972.  A theory of biological pattern
formation.  Kybernet 12, 30-39

\noindent Gray, P., Scott, S. K., 1984. Autocatalytic reactions in the
isothermal continuous stirred tank reactor: oscillations and
instabilities in the system a + 2b $\to$ 3b; b $\to$
c. Chem. Eng. Sci. 39,1087-1097.

\noindent Houchmandzadeh, B., Wieschais, E., Leibler, S., 2002.  
Establishment of developmental precision and proportions in the early
Drosophila embryo.  Nature 415, 798-802

\noindent Hunding, A., S\o rensen, P. G., 1988. Size adaptation of Turing
prepatterns.  J. Math. Biol. 26, 27-39

\noindent Kaneko, K., Yomo, Y., 1994. Cell Division, Differentiation, and
Dynamic Clustering.  Physica D 75, 89-102

\noindent Kaneko, K., Yomo, T., 1999. Isologous Diversification for Robust
Development of Cell Society. J. Theor. Biol. 199, 243-256

\noindent Kondo, S., Asai, R., 1995.  A reaction-diffusion wave on the
skin of the marine angelfish Pomacanthus. Nature 376, 765-768.

\noindent Meinhardt, H., 1982. Models of Biological Pattern Formation. 
Academic Press New York

\noindent Meinhardt, H., Gierer, A., 2000.  Pattern formation by local
self-activation and lateral inhibition BioEssays 22, 753-760.

\noindent Mizuguchi, T., Sano, M., 1995.  Proportion regulation of Biological
Cells in Globally Coupled Nonlinear Systems. Phys. Rev. Lett. 75,
966-969.

\noindent Murray, J. D., 1993.  Mathematical Biology (2nd ed.) 
Spriger-Verlag

\noindent Nicolis, G., Prigogine, I., 1977. Self Organization in 
Non-Equilibrium Systems. J. Wiley and Sons, New York

\noindent Othmer, H. G., Pate, E., 1980. Scale-invariance in 
reaction-diffusion models of spatial pattern
formation.~Proc.~Natl.~Acad.~Sci. 77, 4180-4184

\noindent Ouyang, Q. and Swinney, H. L., 1991
Transition from a uniform state to hexagonal and striped Turing patterns.
Nature 352, 610-611 

\noindent Potter, C. J. and Xu, T., 2001. Mechanism of size control.
Curr. Opin. Gen. Dev. 11. 279-286

\noindent Palmiter, R.D., Brinster, R. L., Hammer, R.E., Trumbauer, M.E.,
Rosenfeld, M.G., Birnberg, N.C., Evans. R.M., 1982. Dramatic growth of
mice that develop from eggs microinjected with metallothionein-growth
hormone fusion genes. Nature 300,611-615.

\noindent Pate, E., Othmer, H. G., 1984. Application of a model for
scale-invariant pattern formation on developing
systems. Differentiation~28, 1-8.

\noindent Pearson, J. E., 1993.  
Complex Patterns in a Simple System. Science 261, 189-192

\noindent Prigogine, I., Lefever, R., 1968.  Symmetry Breaking 
Instabilities in Dissipative Systems II.  J. Chem. Phys. 48,
1695-1700.

\noindent Saunders, P. T., Ho, M. W., 1995. Reliable segmentation by 
successive bifurcation.  Bull. Math. Biol. 57, 539-556.

\noindent Turing, A. M., 1952.  The chemical basis of Morphogenesis.
Philos. Trans. Roy. Soc. Lond. B. 237, 37-72.

\noindent Wolpert, L., 1969. Positional information and the spatial 
pattern of cellar differentiation. J. Theor. Biol. 25, 1-47.

\newpage 

\begin{figure}[tbhp]
\centering
\includegraphics[width=.8\textwidth]{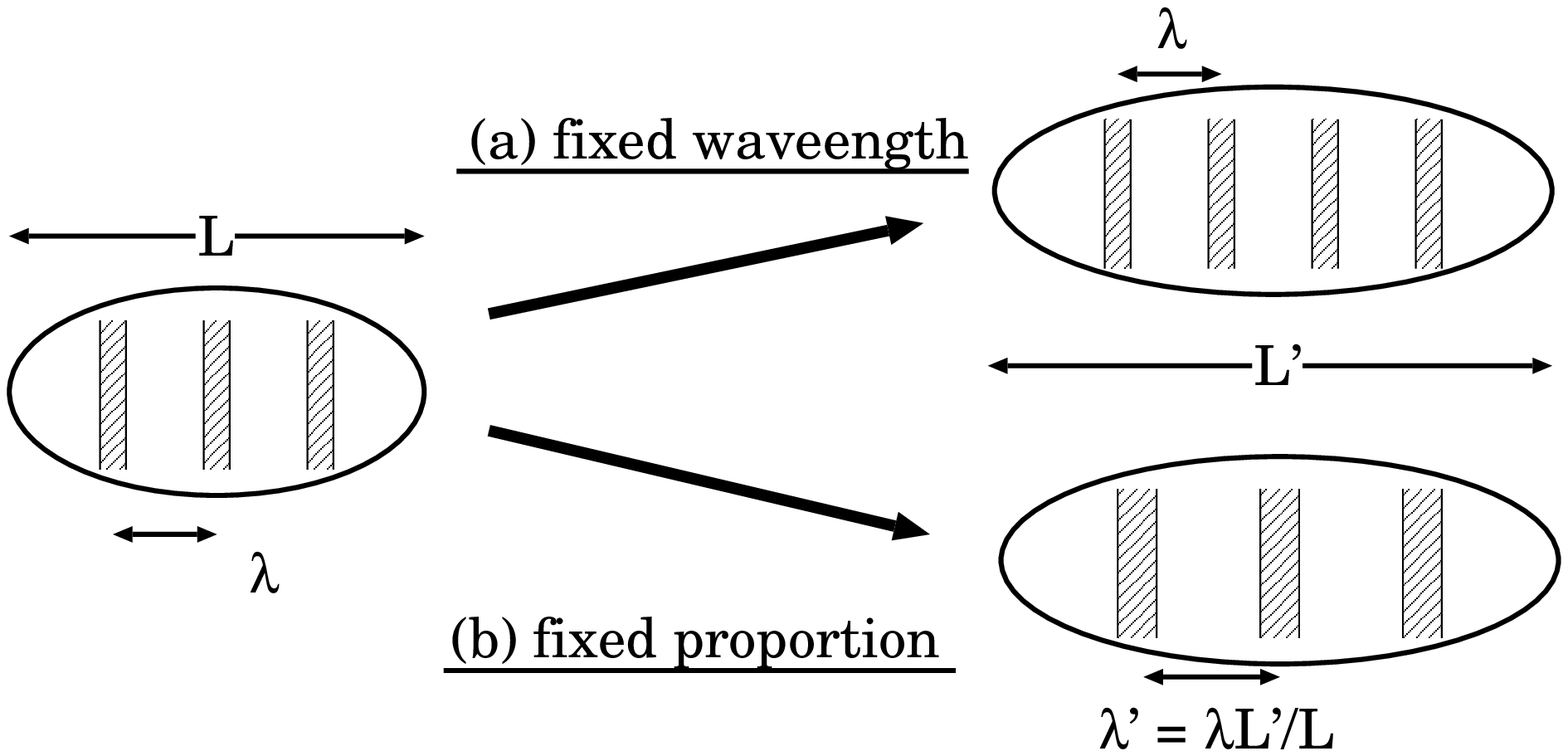}
\caption{Ishihara and Kaneko}
\label{fig:Prop}
\end{figure}

\noindent Figure \ref{fig:Prop} : Pattern formation (a) with a fixed wave 
length and (b) with a fixed proportion.  Ordinary Turing pattern
belongs to the class (a), as the pattern arises by instability of a
certain range of wave lengths.

\begin{figure}[tbhp]
\centering
\includegraphics[width=1.\textwidth]{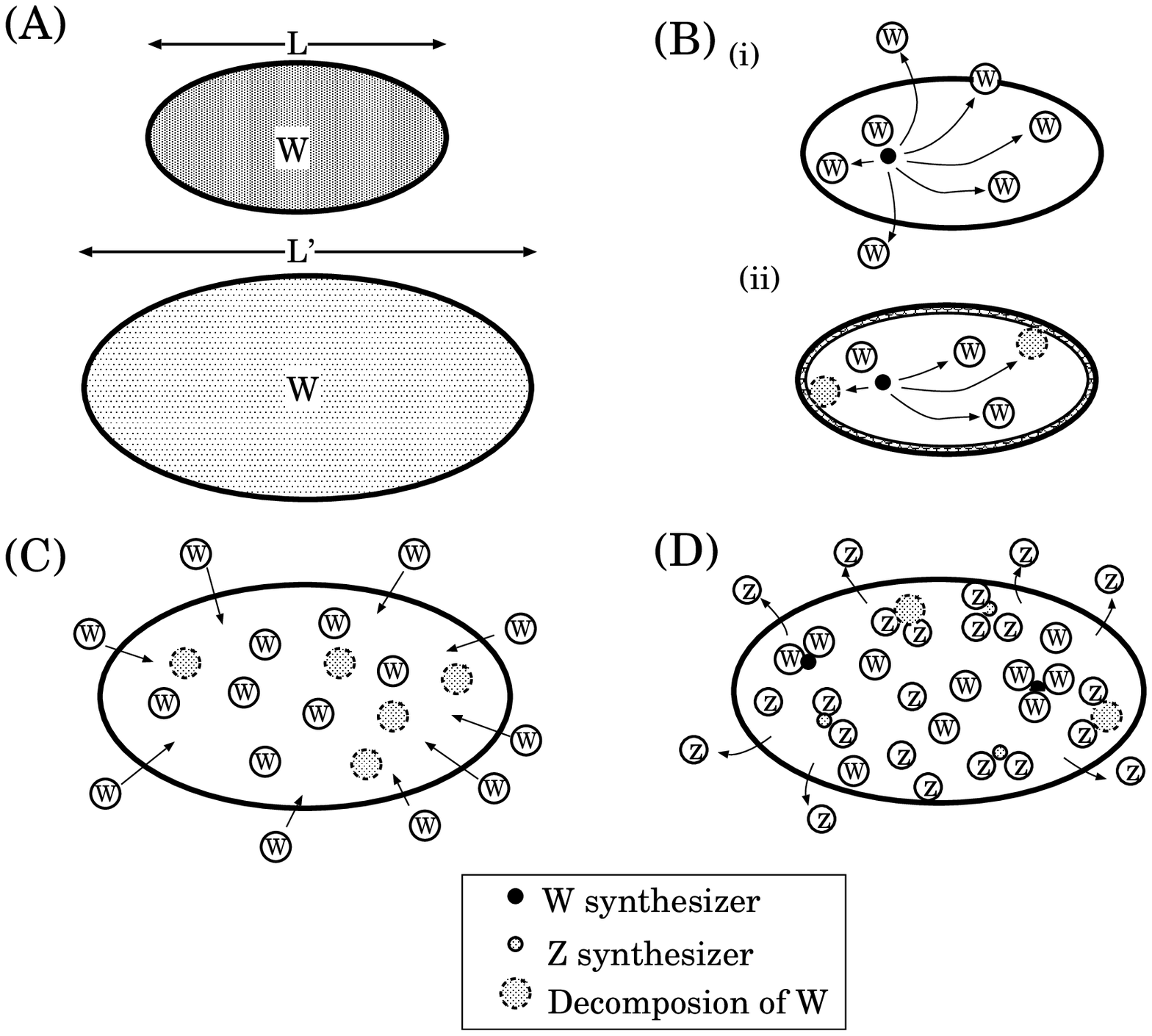}
\caption{Ishihara and Kaneko}
\label{fig:turingcase}
\end{figure}

\noindent Figure \ref{fig:turingcase} : Several cases for the scaling 
behavior of the concentration of W against the system size, as
discussed in the text.

\begin{figure}[tbhp]
\centering
\includegraphics[width=1.\textwidth]{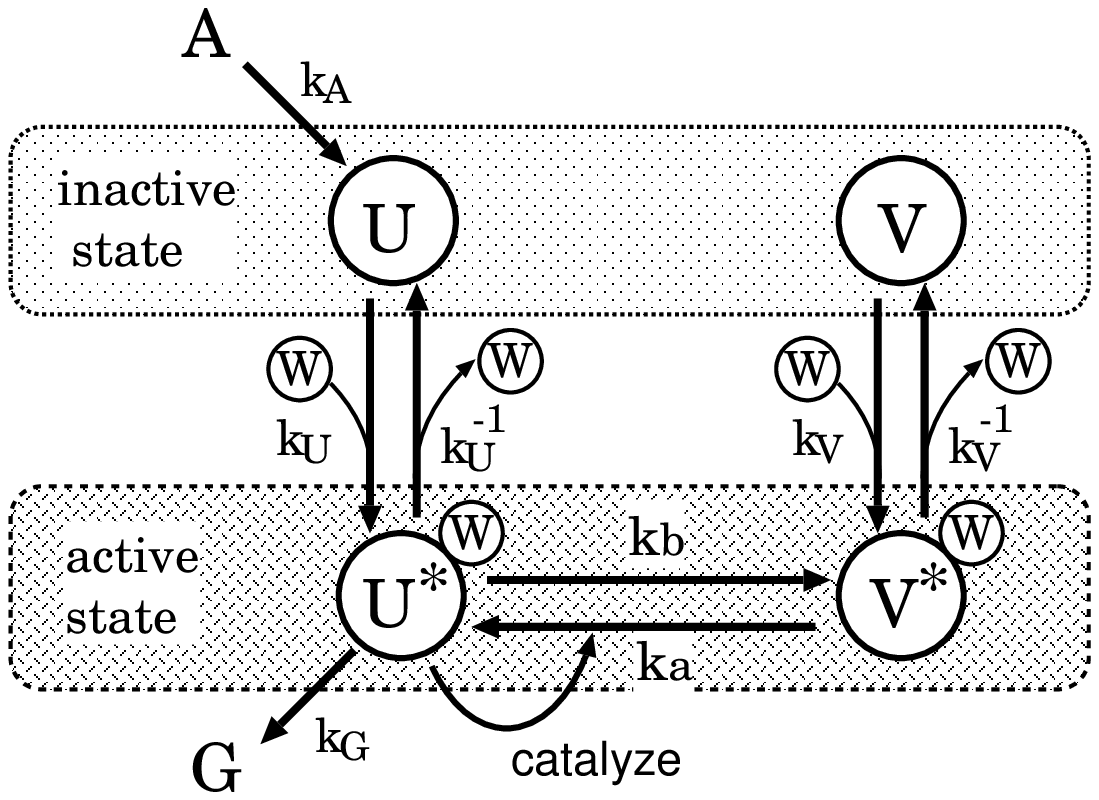}
\caption{Ishihara and Kaneko}
\label{fig:Brusselator}
\end{figure}

\noindent Figure \ref{fig:Brusselator} : Two-state Brusselator model. 
Only chemical components in the active state can react in the model.

\begin{figure}[tbhp]
\centering
\includegraphics[width=1.\textwidth]{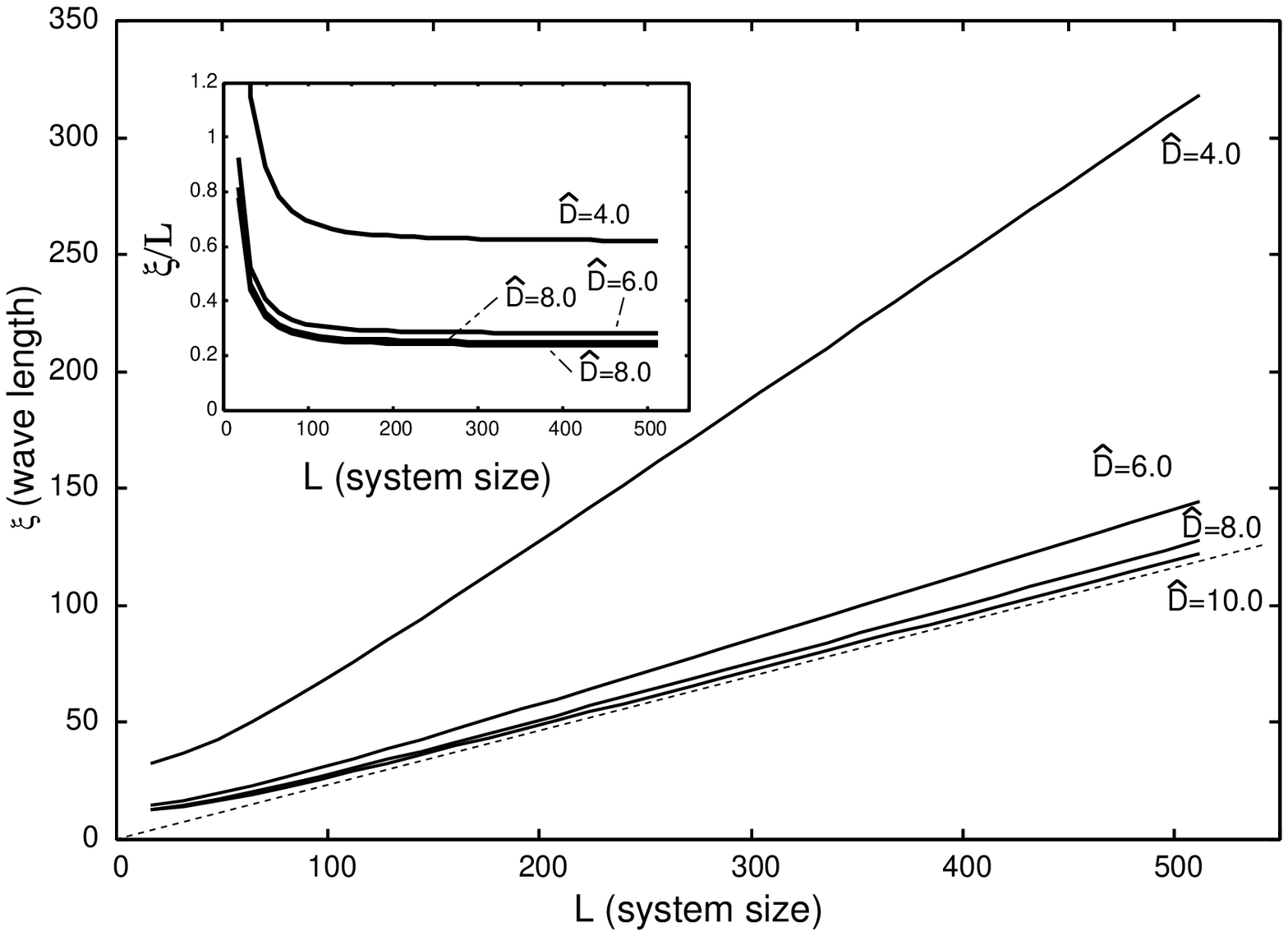}
\caption{Ishihara and Kaneko}
\label{fig:model1x-L}
\end{figure}

\noindent Figure \ref{fig:model1x-L} : Characteristic length $\xi$ of 
the pattern in the two-state Brusselator model (Eq.~(\ref{eq:simueq}))
plotted against the system size $L$. The parameters are $k_U=0.5,
k_U^{-1}=10.0, k_V=0.1, k_V^{-1}=2.0, A=2.0, B=4.0, W_0=5.0 \times
10^4$. We plot for $\hat{D} \equiv D_v/D_u = 4.0, 6.0 8.0,
10.0$. Normalized wavelengths by the system size ($\xi/L$) are plotted
in the inset.

\begin{figure}[tbhp]
\centering
\includegraphics[width=1.\textwidth]{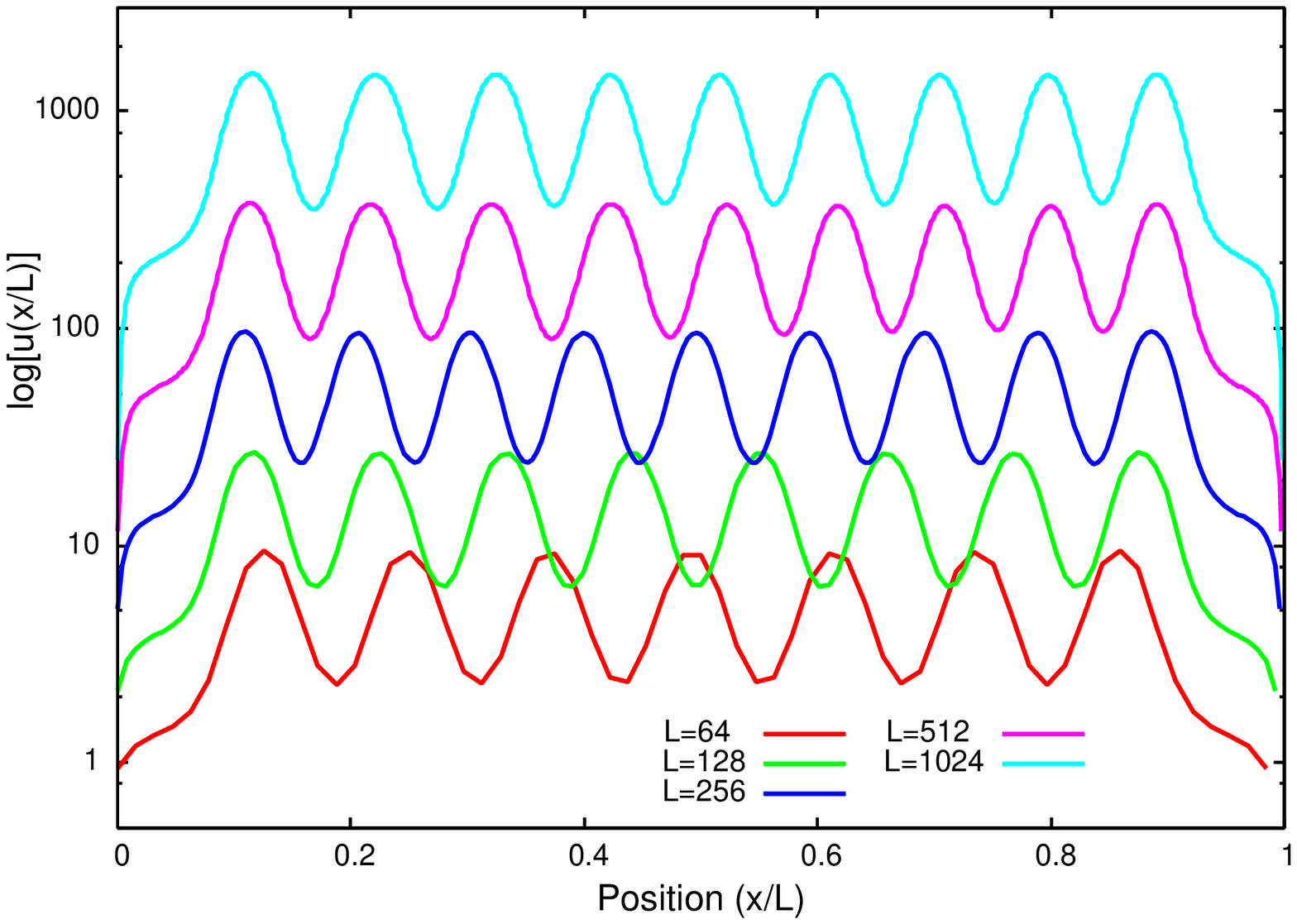}
\caption{Ishihara and Kaneko}
\label{fig:SimulationTuring}
\end{figure}

\noindent Figure \ref{fig:SimulationTuring} : Pattern of the two-state
Brusselator model in Eq.~(\ref{eq:simueq}) obtained numerically, for
the system size $L=$ 64, 128, 256, 512, and 1024. The parameters are
$k_U=0.5, k_U^{-1}=10.0, k_V=0.1, k_V^{-1}=2.0, A=2.0, B=4.0, D_U=0.5,
D_V=3.0, W_0=5.0 \times 10^4$. The number of stripes are invariant
over a wide range of system size, while for too small system size, the
proportion is no longer sustained due to the nonlinearity (saturation)
in $m(w)$ and $n(w)$. Simulations are carried out with the grid size
1.

\begin{figure}[tbhp]
\centering
\includegraphics[width=1.\textwidth]{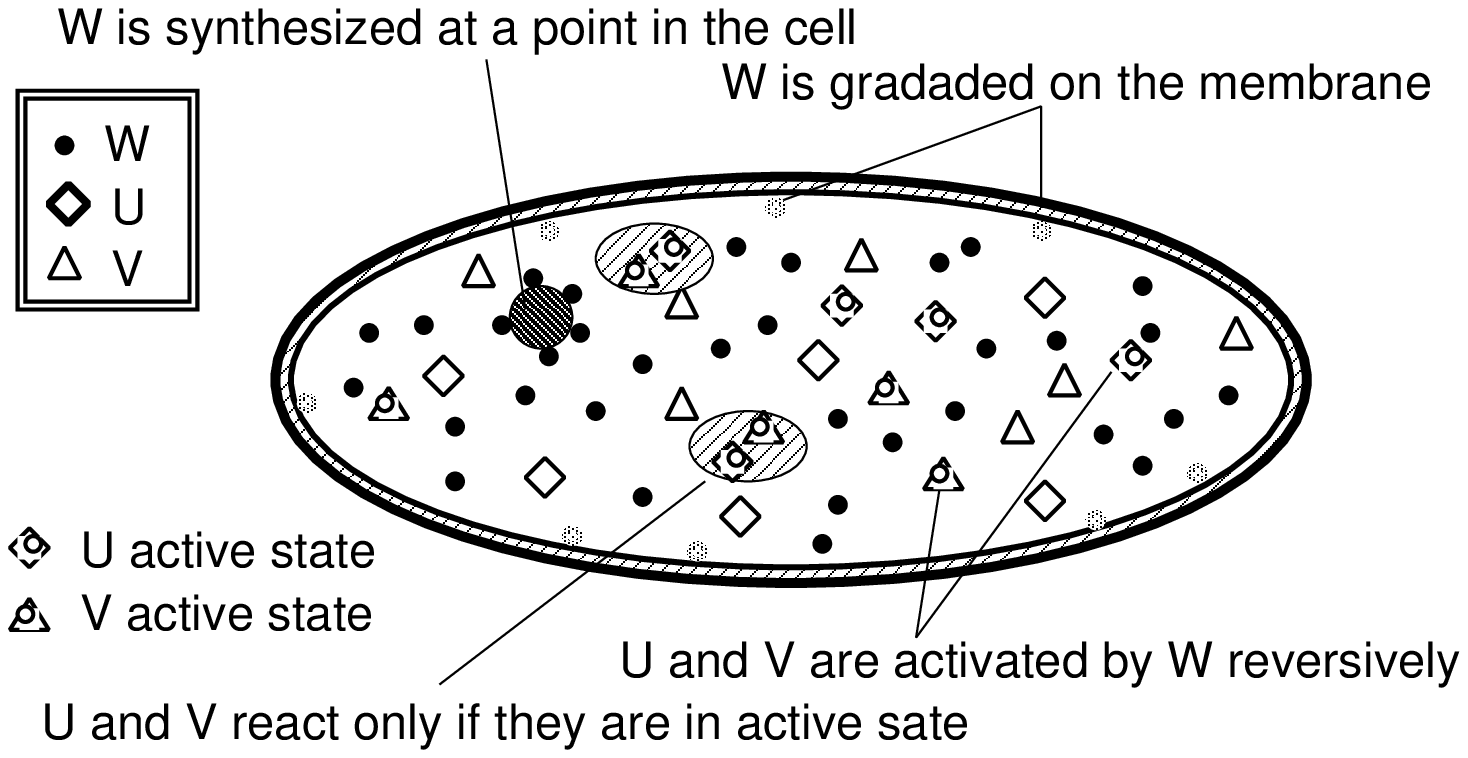}
\caption{Ishihara and Kaneko}
\label{fig:TuringMech}
\end{figure}

\noindent Figure \ref{fig:TuringMech} : A scenario for proportion 
preservation of a Turing pattern.

\begin{figure}[tbhp]
\centering
\includegraphics[width=.6\textwidth]{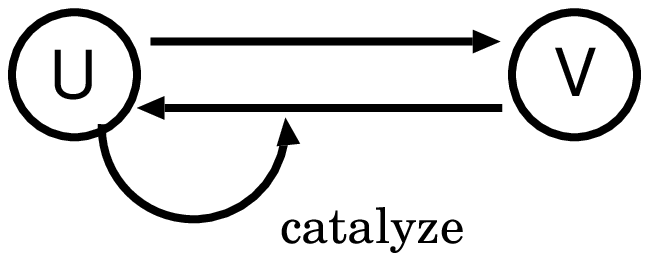}
\caption{Ishihara and Kaneko}
\label{Fig:Model2}
\end{figure}

\noindent Figure \ref{Fig:Model2} : A model with two states of morphogen 
U and V whose sum is conserved, where reaction between them to
regulate the concentration brings about Turing instability.

\begin{figure}[tbhp]
\centering
\includegraphics[width=.9\textwidth]{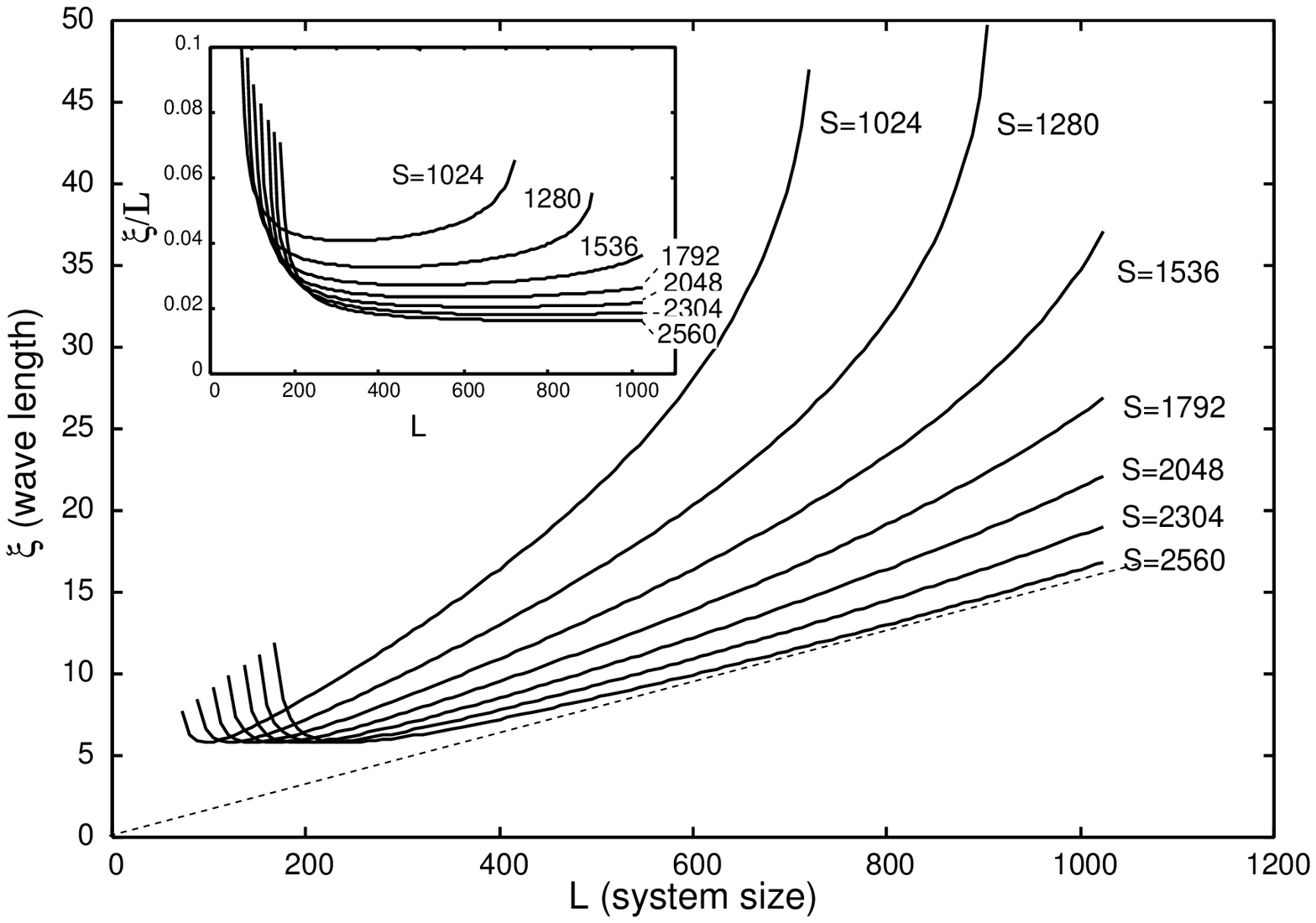}
\caption{Ishihara and Kaneko}
\label{fig:model2anal}
\end{figure}

\noindent Figure \ref{fig:model2anal} : Characteristic length $\xi$ of
the pattern by Eq.~\ref{eq:simueq2} plotted against the system size
$L$, for various values of the total quantity $S\,(= \int
u(x)+v(x)dx)$.  For large $S$, $\xi$ increase in proportion to the
system size $L$.  $B=0.5, \hat{D}=5.0\times 10^2$. Normalized
wavelengths by the system size ($\xi/L$) are plotted in the inset.
 
\begin{figure}[tbhp]
\centering
\includegraphics[width=1.\textwidth]{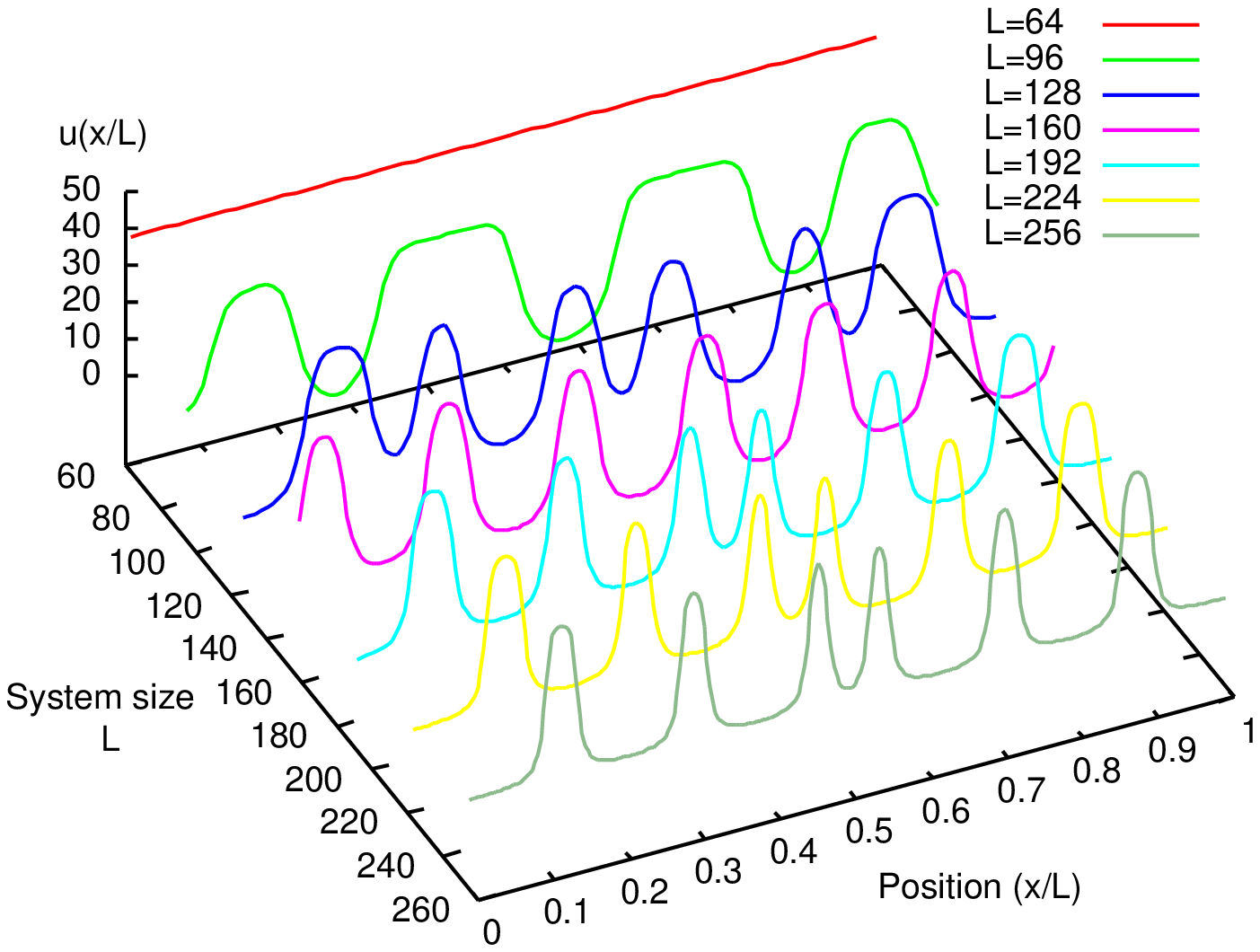}
\caption{Ishihara and Kaneko}
\label{fig:SimulationTuring2}
\end{figure}

\noindent Figure \ref{fig:SimulationTuring2} : Simulation results of 
the modified Brusselator model with a conserved quantity in
Eq.~(\ref{eq:simueq2}) for various size $L=$64, 96, 128, 160, 192, 224
and 256.  The number of stripes are $6$ and invariant for $L \ge 128$.
The parameters are $D_u=1.0 \times 10^2 , D_v=1.0 \times 10^4, B=0.5$,
$S=2560.0$.

\begin{figure}[tbhp]
\centering
\includegraphics[width=.6\textwidth]{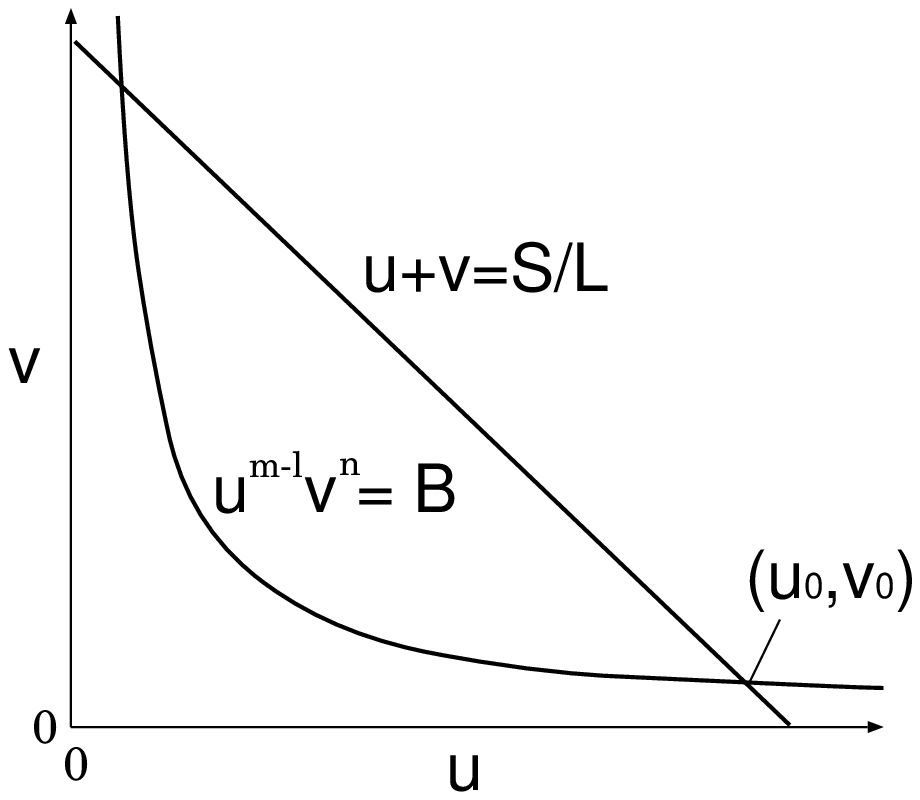}
\caption{Ishihara and Kaneko}
\label{fig:uvApend1}
\end{figure}

\noindent Figure \ref{fig:uvApend1} : Steady state solution plotted in 
$u$-$v$ plane.  The crossing points of the line and the curve give
homogeneous steady solutions of reaction equations
Eq.~(\ref{eq:uvsteady}).


\newpage 

\end{document}